\documentclass[aps,prd,reprint,nofootinbib,superscriptaddress]{revtex4-2}

\usepackage{amsmath,amssymb,bm}
\usepackage{physics}
\usepackage{microtype}
\usepackage{hyperref}

\hypersetup{colorlinks=true,linkcolor=blue,citecolor=blue,urlcolor=blue}

\newcommand{\cA}{\mathcal A}
\newcommand{\tcA}{\widetilde{\mathcal A}}
\newcommand{\tA}{\widetilde A}
\newcommand{\tF}{\widetilde F}
\newcommand{\phie}{\phi_e}
\newcommand{\phim}{\phi_m}
\newcommand{\De}{D^{(e)}}
\newcommand{\Dm}{D^{(m)}}

\begin{document}

\title{Dual symmetry breaking and magnetic charge screening}

\author{Saulo Carneiro}
\thanks{saulocarneiro@on.br\\ORCID: 0000-0001-7098-383X}
\affiliation{Observat\'orio Nacional, 20921-400 Rio de Janeiro, RJ, Brazil}

\begin{abstract}
We extend a Lagrangian formulation of dual electrodynamics by introducing electric and magnetic Higgs scalar fields. The scalars couple to Mandelstam nonlocal potentials which enter the electric and magnetic sectors. We consider the vacuum in which only the magnetic scalar has a nonzero expectation value. The generalised Maxwell equations remain local even though the scalar couplings are nonlocal. In this phase the local electric potential keeps a massless pole while the magnetic sector acquires a mass. The electric interaction is therefore Coulombian at large distances while the pole-pole and charge-pole interactions become short ranged. The magnetic charge is then screened in the sense of the Swieca theorem while the electric charge is not. The Dirac condition is inherited from the symmetric phase as a quantisation rule for the microscopic electric and magnetic couplings. We also comment on screened monopoles as possible dark sector degrees of freedom.
\end{abstract}

\maketitle

\section{Introduction}

The introduction of magnetic charge makes Maxwell electrodynamics manifestly dual at the level of the field equations. The Lagrangian description is less direct because a regular four-potential cannot describe electric and magnetic sources everywhere in space-time. Dirac solved this problem by introducing a singular string \cite{Dirac1931}. Other formulations use two local potentials \cite{CabibboFerrari1962} or a local Lagrangian with a fixed auxiliary $4$-vector \cite{Zwanziger1971}. In general, local formulations are unable to derive the complete field equations without resorting to auxiliary conditions or artifices.

In Ref.~\cite{MelloCarneiroNemes1996} a different route was followed. Two regular local potentials $A_\mu$ and $\tA_\mu$ were introduced to define nonlocal potentials $\cA_\mu$ and $\tcA_\mu$. The action obeys a saddle-point variational principle and gives the complete set of local Maxwell and Lorentz equations with electric and magnetic currents. The same formalism was later quantised and used to construct charge-monopole amplitudes without strings \cite{Carneiro1998}.

Here we add an electric and a magnetic Higgs field \cite{Higgs1964} to this formulation. The scalar fields couple to $\cA_\mu$ and $\tcA_\mu$ rather than directly to the local potentials. We consider a phase in which the magnetic scalar condenses and the electric scalar does not. The resulting breaking of the dual symmetry makes the magnetic sector massive while preserving the massless electric one. The long-distance magnetic charge is then screened. This provides a direct realisation of the relation between a massive Abelian vector sector and charge screening established by Swieca \cite{Swieca1976} and later sharpened by Buchholz and Fredenhagen \cite{BuchholzFredenhagen1979}.

Mass generation has also been considered in local two-photon formulations \cite{Singleton1995,ScottEtAl2018}. The construction below differs at its starting point. The action is the nonlocal dual action of Ref.~\cite{MelloCarneiroNemes1996} and the Higgs fields couple to the nonlocal potentials required by that action. The local equations of motion follow from this nonlocal coupling.

Recent developments have renewed interest in both dual formulations of electrodynamics and the phenomenology of magnetic monopoles. String-free formulations based on dual gauge potentials continue to be investigated \cite{Govaerts2023}, while charge screening in Abelian Higgs models has been analysed in detail in more recent studies \cite{ForgacsLukacs2021}. At the same time, theoretical and experimental aspects of magnetic monopoles remain an active subject, with current reviews covering their foundations and the continuing searches at colliders and in cosmic-ray experiments \cite{Mitsou2026}. The construction considered here provides a different realisation of these ideas, based on the nonlocal action and on spontaneous symmetry breaking in only one of its dual sectors.

We use $\hbar=c=1$, $g_{\mu\nu}=\mathrm{diag}(+,-,-,-)$ and $\square\equiv\partial_\mu\partial^\mu$.

\section{Non-local dual action}

The generalised field tensor is
\begin{equation}
F^{\mu\nu}=\partial^\mu A^\nu-\partial^\nu A^\mu
-\epsilon^{\mu\nu\alpha\beta}\partial_\alpha \tA_\beta,
\label{Fdef}
\end{equation}
with
\begin{equation}
\tF^{\mu\nu}=\frac12\epsilon^{\mu\nu\alpha\beta}F_{\alpha\beta}.
\label{Ftdef}
\end{equation}
The nonlocal potentials are \cite{MelloCarneiroNemes1996}
\begin{equation}
\cA^\mu(x)=A^\mu(x)+\frac12\epsilon^{\mu\gamma\alpha\beta}
\int_P^x \partial_\alpha\tA_\beta(\xi)\,d\xi_\gamma
\label{calA}
\end{equation}
and
\begin{equation}
\tcA^\mu(x)=\tA^\mu(x)-\frac12\epsilon^{\mu\gamma\alpha\beta}
\int_{\widetilde P}^x \partial_\alpha A_\beta(\xi)\,d\xi_\gamma,
\label{caltA}
\end{equation}
where the choice of the integration paths corresponds to a choice of gauge \cite{Carneiro1998}.
They satisfy
\begin{equation}
F^{\mu\nu}=\partial^\mu\cA^\nu-\partial^\nu\cA^\mu
\end{equation}
and
\begin{equation}
\tF^{\mu\nu}=\partial^\mu\tcA^\nu-\partial^\nu\tcA^\mu.
\end{equation}
These relations do not imply homogeneous Maxwell equations. The nonlocal potentials are not regular and their derivatives do not commute,
\begin{equation}
[\partial^\mu,\partial^\nu]\cA^\alpha\ne0,
\end{equation}
with an analogous relation for $\tcA^\alpha$. The local potentials $A_\mu$ and $\tA_\mu$ are regular.

We introduce two complex scalar fields $\phie$ and $\phim$ with covariant derivatives
\begin{equation}
\De_\mu\phie=\left(\partial_\mu+i q_e\cA_\mu\right)\phie
\label{De}
\end{equation}
and
\begin{equation}
\Dm_\mu\phim=\left(\partial_\mu+i q_m\tcA_\mu\right)\phim.
\label{Dm}
\end{equation}
The scalar contribution is written in the saddle form
\begin{equation}
\begin{aligned}
\mathcal L_H={}&
(\De_\mu\phie)^*D^{(e)\mu}\phie-V_e(\phie)\\
&-(\Dm_\mu\phim)^*D^{(m)\mu}\phim+V_m(\phim).
\end{aligned}
\label{LH}
\end{equation}
The relative sign follows the same electric-magnetic saddle structure as the particle action. The opposite overall sign of the dual sector is necessary for obtaining the correct set of equations of motion, that is, the Maxwell equations with electric and magnetic sources, the classical Lorentz equations for the charge and pole and the Klein-Gordon equations for the scalar fields, and it does not spoil the Hamiltonian positivity \cite{MelloCarneiroNemes1996,Carneiro1998}.

Including ordinary electric and magnetic matter currents the action has the form
\begin{equation}
\mathcal L=\mathcal L_0^e+\mathcal L_0^m
-\frac14F_{\mu\nu}F^{\mu\nu}
-j_\mu\cA^\mu+g_\mu\tcA^\mu+\mathcal L_H.
\label{fullL}
\end{equation}
We assume that the scalar sector admits two dual vacuum branches and choose the magnetic one,
\begin{equation}
\langle\phie\rangle=0
\label{vacuum'}
\end{equation}
and
\begin{equation}
\langle\phim\rangle=\frac{v}{\sqrt2}.
\label{vacuum}
\end{equation}
The detailed form of the scalar potential is not needed for the vector spectrum discussed below.

\section{Broken phase and field equations}

We write the magnetic scalar as
\begin{equation}
\phim(x)=\frac{1}{\sqrt2}[v+h(x)]
\exp\left[i\theta(x)/v\right].
\end{equation}
Its kinetic term contains
\begin{equation}
\frac12[v+h(x)]^2
\left[\frac{1}{v}\partial_\mu\theta+q_m\tcA_\mu\right]^2.
\end{equation}
In the unitary gauge $\theta=0$ the quadratic part therefore contains
\begin{equation}
\frac12 m_M^2\tcA_\mu\tcA^\mu,
\label{massterm}
\end{equation}
up to the overall sign of the magnetic sector, with
\begin{equation}
m_M=q_m v.
\label{mass}
\end{equation}
There is no corresponding electric mass term because $\langle\phie\rangle=0$.

The scalar currents obtained from Eq.~\eqref{LH} are
\begin{equation}
j_H^\mu=i q_e\left[\phie^*D^{(e)\mu}\phie-(D^{(e)\mu}\phie)^*\phie\right]
\end{equation}
and
\begin{equation}
g_H^\mu=i q_m\left[\phim^*D^{(m)\mu}\phim-(D^{(m)\mu}\phim)^*\phim\right].
\end{equation}
In the magnetic vacuum
\begin{equation}
g_H^\mu=-m_M^2\tcA^\mu+O(h).
\label{Hcurrent}
\end{equation}
The variation of the nonlocal potentials then gives the local equations \cite{MelloCarneiroNemes1996}\footnote{The variation with respect to the local potentials gives the same result \cite{Carneiro1998}.}
\begin{equation}
\partial_\nu F^{\mu\nu}=-(j^\mu+j_H^\mu)
\label{Maxe}
\end{equation}
and
\begin{equation}
\partial_\nu\tF^{\mu\nu}=-(g^\mu+g_H^\mu).
\label{Maxm}
\end{equation}
To quadratic order around the vacuum, $j_H^\mu$ contains no vector mass term while $g_H^\mu$ is given by Eq.~\eqref{Hcurrent}. It is also straightforward to derive the correct Klein-Gordon equation for the excitation $h(x)$,
\begin{equation}
(\square+m_h^2)h=0,\qquad m_h^2\equiv\left.\frac{\partial^2 V_m}{\partial h^2}\right|_{h=0}.
\label{hKG}
\end{equation}

The regularity of the local potentials is now useful. Substitution of Eq.~\eqref{Fdef} into Eq.~\eqref{Maxe} gives
\begin{equation}
\square A^\mu-\partial^\mu(\partial\cdot A)=j^\mu
\label{Aeom}
\end{equation}
at the quadratic level, because
\begin{equation}
\epsilon^{\mu\nu\alpha\beta}\partial_\nu\partial_\alpha\tA_\beta=0.
\end{equation}
In Lorenz gauge,
\begin{equation}
\square A^\mu=j^\mu.
\label{Awave}
\end{equation}
Thus the electric local potential keeps a massless pole.
For the magnetic potential, on the other hand, one obtains
\begin{equation}
\square\tA^\mu-\partial^\mu(\partial\cdot\tA)
=g^\mu-m_M^2\tcA^\mu.
\label{tAeom}
\end{equation}
The mass is carried by the nonlocal magnetic potential because this is the quantity which enters the scalar covariant derivative.

\section{Gauge freedom after the symmetry breaking}

Before fixing the scalar phases the two nonlocal potentials have their corresponding Abelian gauge freedoms
\begin{equation}
\cA_\mu\rightarrow\cA_\mu+\partial_\mu\Lambda_e
\end{equation}
and
\begin{equation}
\tcA_\mu\rightarrow\tcA_\mu+\partial_\mu\Lambda_m.
\end{equation}
The scalar fields transform with the compensating phases. The magnetic unitary gauge uses the magnetic transformation to set $\theta=0$. The electric gauge freedom remains unbroken because the electric scalar has zero vacuum expectation value.

In addition to the Abelian transformations above, the formalism possesses a generalised gauge freedom of the local potentials \cite{Carneiro1998}. For the source-free quadratic theory relevant for propagator derivations,
this generalised gauge invariance admits a transformation
that sets $A_\mu=0$ while leaving $\tilde{\mathcal A}_\mu$ unchanged.
This transformation therefore preserves the magnetic unitary gauge. In that gauge,
\begin{equation}
\tcA_\mu=\tA_\mu.
\label{Azerorelation}
\end{equation}
Equation~\eqref{tAeom} then becomes, in Lorenz gauge,
\begin{equation}
(\square+m_M^2)\tA^\mu=g^\mu.
\label{Proca}
\end{equation}
The dual condition $\tA_\mu=0$ is no longer valid as an independent gauge choice after the magnetic unitary gauge has been imposed. This is the asymmetry produced by the vacuum choice in Eq.~\eqref{vacuum}.

\section{Propagators and long-distance interactions}

The charge-charge, pole-pole and charge-pole propagators are given by the time-ordered $\langle0|{\cal A}{\cal A}|0\rangle$, $\langle0|\tilde{\cal A}\tilde{\cal A}|0\rangle$ and $\langle0|{\cal A}\tilde{\cal A}|0\rangle$, respectively \cite{Carneiro1998}.
The massless contribution to the electric propagator follows directly from Eq.~\eqref{Awave}. In the same normalisation used in Ref.~\cite{Carneiro1998} and after contraction with a conserved current it is
\begin{equation}
D_{\mu\nu}(k)=\frac{4\pi}{k^2+i\epsilon}g_{\mu\nu}.
\label{Dmassless}
\end{equation}

The magnetic propagator follows from Eq.~\eqref{Proca}. Longitudinal terms do not contribute between conserved magnetic currents and we can write
\begin{equation}
\widetilde D_{\mu\nu}(k)=
\frac{4\pi}{k^2-m_M^2+i\epsilon}g_{\mu\nu}.
\label{Dmassive}
\end{equation}
The pole-pole amplitude therefore has a Yukawa structure. In the static limit,
\begin{equation}
V_{mm}(r)\propto g_1g_2\frac{e^{-m_Mr}}{r}.
\label{Vmm}
\end{equation}

The mixed propagator of Ref.~\cite{Carneiro1998}, which intermediates the charge-pole interaction, can be calculated in the same gauge $A_\mu=0$. Its Eq.~(17) gives
\begin{equation}
C_{\mu\nu}(x)=
\epsilon_{\mu\gamma\alpha\beta}
\int_P^x d\xi^\gamma\,
\partial_\xi^\alpha\widetilde D^\beta{}_{\nu}(\xi).
\label{Cposition}
\end{equation}
Taking its derivative and transforming to momentum space we have
\begin{equation}
k_\lambda C_{\mu\nu}(k)=
\epsilon_{\mu\lambda\alpha\beta}k^\alpha
\widetilde D^\beta{}_{\nu}(k),
\end{equation}
and then
\begin{equation}
k_\lambda C_{\mu\nu}(k)=
\frac{4\pi}{k^2-m_M^2+i\epsilon}
 k^\alpha\epsilon_{\mu\nu\lambda\alpha}.
\label{Cmassive}
\end{equation}
The massless pole which appears in the symmetric phase is absent. The charge-pole interaction is therefore short ranged as well.

For charge-charge scattering the local electric potential cannot acquire a Proca term because Eq.~\eqref{Aeom} is unchanged at quadratic order. For a purely electric source Eq.~\eqref{Maxe} is the ordinary Maxwell equation and gives the Coulomb field. The propagator of $\cA_\mu$ may also contain fluctuations involving the massive magnetic potential. Such terms have the magnetic scale $m_M$ and do not generate an additional physical pole at $k^2=0$. The resulting infrared propagator is consequently given by (\ref{Dmassless}) plus terms with no physical pole at $k^2=0$. In coordinate space,
\begin{equation}
V_{ee}(r)=\frac{e_1e_2}{r}+V_{\rm short}(r).
\label{Vee}
\end{equation}
The first term is the ordinary Coulomb interaction. The second term is governed by the magnetic mass scale and vanishes at large distances.

The infrared content of the broken phase can then be summarised as
\begin{equation} \nonumber
 e-e\quad\longrightarrow\quad \text{long range},
\end{equation}
\begin{equation} \nonumber
 g-g\quad\longrightarrow\quad \text{short range},
\end{equation}
\begin{equation} \nonumber
 e-g\quad\longrightarrow\quad \text{short range}.
\end{equation}

\section{Magnetic screening and the Swieca theorem}

The result above has a direct interpretation in terms of the Maxwell charge. Define the magnetic current by
\begin{equation}
j_{M,m}^\mu=-\partial_\nu\tF^{\mu\nu}.
\label{magmaxcurrent}
\end{equation}
The associated charge can be written as a surface integral
\begin{equation}
Q_{M,m}=\int d^3x\,j_{M,m}^0
=-\lim_{R\rightarrow\infty}
\oint_{S_R}dS_i\,\tF^{0i}.
\label{magcharge}
\end{equation}
The magnetic propagator has no massless pole after the breaking. A field generated by a localised magnetic source therefore decreases exponentially on the scale $m_M^{-1}$. Equation~\eqref{magcharge} then gives
\begin{equation}
Q_{M,m}=0.
\label{screened}
\end{equation}

This is the magnetic version of the screening result established by Swieca \cite{Swieca1976}. The later analysis of Buchholz and Fredenhagen made this connection more precise by relating the existence of nonzero Abelian charge to the presence of massless excitations in the corresponding gauge sector \cite{BuchholzFredenhagen1979}. In the present theory the relevant statement is applied to the magnetic sector. The corresponding propagator has no massless pole and its Maxwell charge is screened.

The microscopic coupling $g$ has not disappeared. A monopole still carries the magnetic charge which enters the local interaction. The condensate produces the compensating long-distance response. The monopole together with this response has zero magnetic Maxwell charge at infinity.

The electric sector behaves differently. Equation~\eqref{Dmassless} contains a massless pole and the electric flux through a sphere at infinity need not vanish. Hence
\begin{equation}
Q_{M,e}\ne0
\end{equation}
is allowed. The broken phase screens magnetic charge without screening electric charge.

\section{Charge quantisation}

Before symmetry breaking the theory is the dual electrodynamics, described in Refs.~\cite{MelloCarneiroNemes1996,Carneiro1998} at the Lagrangian level. Electric and magnetic charges therefore obey the Dirac condition
\begin{equation}
eg=\frac{n}{2},
\label{Diraccondition}
\end{equation}
in the units used here. With $\hbar$ and $c$ restored the right hand side is $n\hbar c/2$. More generally the Schwinger-Zwanziger condition applies to dyons \cite{Schwinger1966,Zwanziger1971}.

The symmetry breaking changes the vacuum and the infrared propagators. It does not change the microscopic couplings which label the representations of the gauge group. The magnetic Higgs field itself must carry a charge $q_m$ belonging to the same charge lattice before it condenses. We therefore keep Eq.~\eqref{Diraccondition} as the quantisation rule for the microscopic monopole charge in the broken phase. Screening changes the asymptotic Maxwell charge and does not redefine the coupling $g$.

A different conclusion was reached in the local two-photon construction of Ref.~\cite{ScottEtAl2018}. There the field angular momentum of a Coulomb electric field and a Yukawa magnetic field gives a modified charge quantisation rule. There the two local potentials are taken as two physical gauge bosons and the quantisation condition is reconstructed from the angular momentum of the broken-phase fields. In the present nonlocal formulation the Dirac lattice already exists in the symmetric phase and the broken phase is obtained from the same theory by changing the vacuum. The microscopic relation in Eq.~\eqref{Diraccondition} is therefore inherited rather than reconstructed from the screened field at finite separation.

\section{Strong–Weak Asymmetry and Vacuum Selection}

The vacuum choice adopted in Eqs.~(\ref{vacuum'})-(\ref{vacuum}) may have a natural quantum motivation. At the classical level the electric and magnetic sectors are related by the dual structure of the theory. This symmetry is no longer reflected in the strength of the corresponding quantum interactions once charge quantisation is imposed. The Dirac condition relates the microscopic electric and magnetic couplings through Eq.~(\ref{Diraccondition}).
Consequently, when the electric coupling is perturbative, the magnetic coupling is necessarily strong. The two sectors may therefore have very different quantum vacuum structures even when their classical descriptions are dual.

This observation suggests a possible dynamical origin for the asymmetric vacuum.
Quantum corrections to the electric scalar sector can remain perturbative, while the magnetic scalar is coupled to a strongly interacting sector. There is then no reason for the quantum effective potentials of $\phi_e$ and $\phi_m$ to retain the symmetry of their classical counterparts. In particular, nonperturbative magnetic dynamics may favour a vacuum with nonzero $\langle\phi_m\rangle$, while the symmetric configuration $\langle\phi_e\rangle=0$ remains stable in the electric sector.

The situation is qualitatively reminiscent of radiatively generated symmetry breaking \cite{ColemanWeinberg1973}, although the magnetic sector cannot in general be treated by an ordinary perturbative effective-potential expansion. Because the magnetic coupling fixed by the Dirac condition is large when the electric coupling is small, the determination of the magnetic effective potential would require a genuinely nonperturbative treatment. The present construction does not attempt such a derivation.

The asymmetric vacuum assumed here should therefore be regarded as a phenomenological realisation of a possibility suggested by the quantum strong--weak asymmetry between the two sectors. If the magnetic condensate is indeed generated dynamically, the breaking of the classical dual symmetry would then be traced to the quantum relation between electric and magnetic charges rather than imposed independently at the level of the vacuum.

\section{Dark sector implications}

The broken phase contains a natural dark sector candidate: the screened monopole. It can be massive and still have no magnetic Coulomb field at distances larger than $m_M^{-1}$. Its electromagnetic interaction with ordinary matter is then short ranged even though its microscopic magnetic charge is not zero. Conventional bounds on monopoles often rely on long-range acceleration by galactic or primordial magnetic fields \cite{KobayashiPerri2023,ZhangEtAl2024}. Such bounds cannot be transferred without modification when the magnetic interaction is screened on scales shorter than those entering the bound. Its viability depends on its lifetime and on the available channels into scalar excitations. A dark matter interpretation requires a cosmologically stable state and a production mechanism giving the observed relic density. These questions depend on the scalar potential and on the mass hierarchy and are beyond the present discussion.

The relevant point here is more limited. Magnetic screening removes the long-range electromagnetic signature which normally makes a Dirac monopole easy to accelerate and difficult to hide. This opens a phenomenological regime which is absent for an unscreened monopole.

\section{Concluding remarks}

We have introduced two Higgs fields in the nonlocal Lagrangian formulation of dual electrodynamics given in Refs.~\cite{MelloCarneiroNemes1996,Carneiro1998}. The electric Higgs couples to the nonlocal potential $\cA_\mu$ and the magnetic Higgs couples to $\tcA_\mu$. We selected the vacuum in which only the magnetic field has a nonzero expectation value.

The variation of the nonlocal action continues to give local field equations. In the broken phase the local electric potential obeys the same massless wave equation as in the symmetric theory. The magnetic equation contains the Higgs current and in the $A_\mu=0$ gauge it becomes a massive vector equation with $m_M=q_mv$. The gauge choice $\tA_\mu=0$, however, is no longer independently available after the magnetic Higgs phase has been removed.

The propagators then give a simple infrared picture. The electric propagator keeps its $k^2=0$ pole. The magnetic propagator has a pole at $k^2=m_M^2$, and the mixed propagator inherits the same massive structure. Electric charge remains Coulombian at large distances while pole-pole and charge-pole interactions are short ranged.

The disappearance of the massless pole in the magnetic and mixed propagators has a structural consequence. The magnetic Maxwell charge vanishes according to the screening theorem of Swieca while the electric Maxwell charge can remain nonzero. The monopole charge $g$ survives as a microscopic quantised coupling and obeys the Dirac condition inherited from the symmetric phase. The quantity which vanishes is the asymptotic magnetic Maxwell charge of the dressed state.

The Dirac condition also means that the microscopic magnetic coupling is intrinsically nonperturbative when the electric coupling is perturbative. The role of the tree-level propagators above is to identify the pole structure and the corresponding infrared range of the interactions. They should not be interpreted as a precise expansion in powers of $g$. This does not affect the screening conclusion. Once the magnetic sector has a mass gap, its exact long-distance correlations are exponentially suppressed and the Swieca result gives a vanishing magnetic Maxwell charge independently of a perturbative expansion in $g$. The effective magnetic interaction therefore tends to zero at distances large compared with $m_M^{-1}$ even though the microscopic Dirac charge remains large and quantised.

This phase therefore realises a dual electrodynamics in which electric charge is visible at long distances and magnetic charge is screened. The mechanism follows from the nonlocal dual action and does not require the introduction of an independent massless magnetic photon in the symmetric phase.

The possibility that monopoles may evade conventional searches remains phenomenologically relevant. Recent collider searches continue to place direct constraints on monopole production \cite{MoEDAL2024}, while alternative mechanisms have been proposed in which heavy monopoles may become difficult to observe through the formation of strongly bound states \cite{Fanchiotti2025}. The screening mechanism discussed here is different, since it suppresses the long-range magnetic interaction itself. It therefore provides another setting in which the absence of an observable long-range monopole field does not necessarily imply the absence of a microscopic magnetic charge.

\section*{Acknowledgements}

I am thankful to CNPq (Brazil) for support through the grant n. 308518/2023-3.


\begin{thebibliography}{99}

\bibitem{Dirac1931}
P. A. M. Dirac, Proc. R. Soc. A \textbf{133}, 60 (1931).

\bibitem{CabibboFerrari1962}
N. Cabibbo and E. Ferrari, Nuovo Cimento \textbf{23}, 1147 (1962).

\bibitem{Zwanziger1971}
D. Zwanziger, Phys. Rev. D \textbf{3}, 880 (1971).

\bibitem{MelloCarneiroNemes1996}
P. C. R. Cardoso de Mello, S. Carneiro and M. C. Nemes, Phys. Lett. B \textbf{384}, 197 (1996).

\bibitem{Carneiro1998}
S. Carneiro, JHEP \textbf{07}, 022 (1998).

\bibitem{Higgs1964}
P. W. Higgs, Phys. Lett. \textbf{12}, 132 (1964).

\bibitem{Swieca1976}
J. A. Swieca, Phys. Rev. D \textbf{13}, 312 (1976).

\bibitem{BuchholzFredenhagen1979}
D. Buchholz and K. Fredenhagen, Nucl. Phys. B \textbf{154}, 226 (1979).

\bibitem{Singleton1995}
D. Singleton, Int. J. Theor. Phys. \textbf{34}, 37 (1995).

\bibitem{ScottEtAl2018}
J. Scott, T. J. Evans, D. Singleton, V. Dzhunushaliev and V. Folomeev, Eur. Phys. J. C \textbf{78}, 382 (2018).

\bibitem{Govaerts2023}
J. Govaerts, Eur. Phys. J. C \textbf{83}, 158 (2023).

\bibitem{ForgacsLukacs2021}
P. Forg\'acs and \'A. Luk\'acs, Eur. Phys. J. C \textbf{81}, 243 (2021).

\bibitem{Mitsou2026}
V. A. Mitsou, Eur. Phys. J. Spec. Top. (2026),
doi:10.1140/epjs/s11734-026-02463-z.

\bibitem{Schwinger1966}
J. Schwinger, Phys. Rev. \textbf{144}, 1087 (1966).

%\bibitem{GoldhaberHeras2017}
%A. S. Goldhaber and R. Heras, arXiv 1710.03321 (2017).

\bibitem{ColemanWeinberg1973} S. Coleman and E. Weinberg, Phys. Rev. D \textbf{7}, 1888 (1973).

\bibitem{KobayashiPerri2023}
T. Kobayashi and D. Perri, Phys. Rev. D \textbf{108}, 083005 (2023).

\bibitem{ZhangEtAl2024}
C. Zhang, S.-H. Zhang, B. Fu, J.-F. Zhang and X. Zhang, JHEP \textbf{08}, 220 (2024).

\bibitem{MoEDAL2024}
B. Acharya et al., Phys. Rev. Lett. \textbf{133}, 071803 (2024).

\bibitem{Fanchiotti2025}
H. Fanchiotti, C. A. Garc\'ia Canal and V. Vento,
Eur. Phys. J. Plus \textbf{140}, 170 (2025).

%\bibitem{ArcadiEtAl2020}
%G. Arcadi, A. Djouadi and M. Kado, Phys. Lett. B \textbf{805}, 135427 (2020).

\end{thebibliography}
\end{document}